\documentclass[10pt,aps,prb,twocolumn,superscriptaddress]{revtex4-2}
\usepackage{amsmath}
\usepackage{fixltx2e}
\usepackage{hyperref}
\hypersetup{colorlinks=true, allcolors=blue}
\usepackage{graphicx}
\usepackage{subfigure}
\usepackage{xspace}

\usepackage[dvipsnames]{xcolor}

\usepackage{amsfonts}
\usepackage{mathrsfs}
\usepackage{dsfont}
\usepackage{mathtools}
\usepackage[capitalize]{cleveref}
\usepackage{simplewick}
\usepackage{hyperref}
\usepackage{float}
\usepackage[normalem]{ulem}

\usepackage{tabularx}
\usepackage{amstext} %
\usepackage{array}   %
\usepackage{multirow}

\newcolumntype{L}{>{$}l<{$}} %
\newcolumntype{C}{>{$}c<{$}} %

\begin{document}

\newcommand\kr[1]{\textcolor{blue}{(KR: #1)}}
\newcommand\dk[1]{\textcolor{magenta}{(DK: #1)}}
\newcommand\rv[1]{\textcolor{orange}{(RV: #1)}}
\renewcommand\sb[1]{\textcolor{red}{(SB: #1)}}
\newcommand\ar[1]{\textcolor{brown}{(AR: #1)}}
\newcommand{\corout}[1]{{\color{black}\sout{#1}}}

\title{Field-induced effects in the spin liquid candidate PbCuTe$_{2}$O$_{6}$}

\author{P. Eibisch}
\affiliation{Physikalisches Institut, Goethe-Universit\"at, 60438 Frankfurt (M), Germany}
\author{C. Thurn}
\affiliation{Physikalisches Institut, Goethe-Universit\"at, 60438 Frankfurt (M), Germany}
\author{A. Ata}
\affiliation{Physikalisches Institut, Goethe-Universit\"at, 60438 Frankfurt (M), Germany}
\author{Y. Saito}
\affiliation{Physikalisches Institut, Goethe-Universit\"at, 60438 Frankfurt (M), Germany}
\author{S. Hartmann}
\affiliation{Physikalisches Institut, Goethe-Universit\"at, 60438 Frankfurt (M), Germany}
\author{U. Tutsch}
\affiliation{Physikalisches Institut, Goethe-Universit\"at, 60438 Frankfurt (M), Germany}
\author{B. Wolf}
\affiliation{Physikalisches Institut, Goethe-Universit\"at, 60438 Frankfurt (M), Germany}
\author{A. T. M. Nazmul Islam}
\affiliation{Helmholtz-Zentrum Berlin f\"ur Materialien und Energie, 14109 Berlin, Germany}
\author{S. Chillal}
\affiliation{Helmholtz-Zentrum Berlin f\"ur Materialien und Energie, 14109 Berlin, Germany}
\author{A. R. N. Hanna}
\affiliation{Helmholtz-Zentrum Berlin f\"ur Materialien und Energie, 14109 Berlin, Germany}
\affiliation{Institut f\"ur Festk\"orperforschung, Technische Universit\"at Berlin, 10623 Berlin, Germany}
\author{B. Lake}
\affiliation{Helmholtz-Zentrum Berlin f\"ur Materialien und Energie, 14109 Berlin, Germany}
\affiliation{Institut f\"ur Festk\"orperforschung, Technische Universit\"at Berlin, 10623 Berlin, Germany}
\author{M. Lang}
\affiliation{Physikalisches Institut, Goethe-Universit\"at, 60438 Frankfurt (M), Germany}

\date{\today}

\begin{abstract}
PbCuTe$_2$O$_6$ is considered as one of the rare candidate materials for a three-dimensional quantum spin liquid (QSL). This assessment was based on the results of various magnetic experiments, performed mainly on polycrystalline material. More recent measurements on single crystals revealed an even more exotic behavior, yielding ferroelectric order below $T_{\text{FE}}\approx 1\,\text{K}$, accompanied by distinct lattice distortions, and a somewhat modified magnetic response which is still consistent with a QSL. Here we report on low-temperature measurements of various thermodynamic, magnetic and dielectric properties of single crystalline PbCuTe$_2$O$_6$ in magnetic fields $B\leq 14.5\,\text{T}$. The combination of these various probes allows us to construct a detailed $B$-$T$ phase diagram including a ferroelectric phase for $B \leq$ 8\,T and a $B$-induced magnetic phase at $B \geq$ 11\,T. These phases are preceded by or coincide with a structural transition from a cubic high-temperature phase into a distorted non-cubic low-temperature state. The phase diagram discloses a ferroelectric quantum critical point at $B_{c1} = 7.9\,\text{T}$, where the second-order phase transition line associated with ferroelectric order is suppressed to zero. In addition, a magnetic quantum phase transition is revealed at $B_{c2} = 11\,\text{T}$. The corresponding phase transition to a field-induced magnetic order at $B > B_{c2}$ is likely to be of first order. Field-induced lattice distortions, observed in the state at $T>$ 1\,K and which are assigned to the effect of spin-orbit interaction of the Cu$^{2+}$-ions, are considered as the key mechanism by which the magnetic field couples to the dielectric degrees of freedom in this material.
\end{abstract}


\maketitle

%-------------------%
%  1. Introduction and motivation %
%-------------------%
\section{Introduction}
%\subsection{}
%\subsubsection{}
The compound PbCuTe$_2$O$_6$ has attracted considerable interest as one of the rare candidates for a three-dimensional (3D) quantum spin liquid (QSL), see Refs.\,\cite{Savary2017, Broholm2020} for recent reviews on QSLs. In this material, which crystallizes in a non-centrosymmetric cubic structure, the Cu$^{2+}$ ($S = 1/2$)  spins are coupled by isotropic antiferromagnetic interactions $J_1, J_2, J_3, J_4 > 0$ \cite{Chillal2020}.  Density functional theory calculations, based on room-temperature structural data, reveal for the first and second-nearest-neighbor interaction $J_1$ = 1.13\,meV and $J_2$ = 1.07\,meV, respectively. Whereas $J_1$ forms isolated equilateral triangles, $J_2$ forms a strongly frustrated 3D network of corner-sharing triangles known as a hyperkagome lattice. The combination of $J_1$ and $J_2$ leads to a special lattice where each Cu$^{2+}$ is involved in three different, corner-sharing triangles \cite{Chillal2020}. The weaker third- and fourth-nearest-neighbor interactions of respectively $J_3$ = 0.59\,meV and $J_4$ = 0.12\,meV couple the spins into chains. See \textbf{Appendix} \ref{AppendixC} for details of the structure, including the magnetic units and their couplings. \\
\indent
Previous magnetic investigations, performed by using pressed-powder or polycrystalline samples, revealed no indications for long-range magnetic order down to temperatures as low as 0.02\,K \cite{Khuntia2016}. In addition, the magnetic spectrum obtained by inelastic neutron scattering experiments revealed diffuse continua, consistent with fractional spinon excitations \cite{Chillal2020} as expected for a QSL. More recently, thanks to the availability of high-quality single crystals of this material \cite{Hanna2021}, the discussion has taken a new twist. It was found that in these single crystals $-$ as opposed to the polycrystalline or powder materials studied until then \cite{Koteswararao2014, Khuntia2016, Chillal2020} - a phase transition into ferroelectric order shows up at $T_{\text{FE}}\approx 1\, \text{K}$ \cite{Thurn2021}, which is accompanied by pronounced lattice distortions. Moreover, a somewhat modified magnetic response was observed, which indicates the absence of magnetic order and is still consistent with a QSL, but with clear indications for quantum critical behavior \cite{Thurn2021}. It was argued in Ref.\,\cite{Thurn2021}, and subsequently demonstrated in Ref.\,\cite{Hanna2021}, that the occurrence of ferroelectric order depends critically on the size of the single crystalline grains in the material, reminiscent of observations made in ferroelectric ceramics, see, e.g., Ref.\,\cite{Zhao2004}. For PbCuTe$_2$O$_6$ these studies revealed that the phase transition anomaly to ferroelectric order becomes drastically reduced for grain sizes of about 30\,$\mu$m and completely vanishes for crystallites with diameter below 10\,$\mu$m \cite{Hanna2021}. This finding provided a rationale for the absence of ferroelectric order in the pressed-powder samples of PbCuTe$_2$O$_6$ studied in Refs.\,\cite{Koteswararao2014, Khuntia2016, Chillal2020}, see Ref.\,\cite{Thurn2021} for details.
\newline\\
\indent
The observation of ferroelectric order accompanied by pronounced lattice distortions and a modified magnetic response in single crystalline PbCuTe$_2$O$_6$ \cite{Thurn2021} highlights a distinct coupling between dielectric, lattice and magnetic degrees of freedom in this material. For a deeper understanding of this coupling and its implications for the material's ground state, we present here a detailed investigation of the low-temperature thermodynamic, magnetic and dielectric properties of single crystalline PbCuTe$_{2}$O$_{6}$ in magnetic fields $B\leq 14.5\,\text{T}$. The focus of our experimental investigations lies on the field orientation parallel to [110]. Along this direction, the anomalies in the coefficient of thermal expansion and in the dielectric constant were found to be significantly stronger as compared to those along the [100] direction (see supplemental information of Ref.\,\cite{Thurn2021}). In fact, the high sensitivity of the [110] axis to changes in the dielectric degrees of freedom can be rationalized by considering the crystal structure \cite{Weil2018}, providing channels along [110], in which the lone pairs of the Pb$^{2+}$ ions may protrude. Our study reveals a rich phase diagram, including a ferroelectric phase at zero to moderately strong magnetic fields and a high-field induced magnetically-ordered phase, both of which are preceded by or coincide with a phase transition which is likely of structural nature. We discuss the character of the various phases and a possible mechanism by which the magnetic field couples to the dielectric degrees of freedom.

\section{Experiment}
The single crystals investigated were prepared by using either a top-seeded-solution-growth (TSSG) method or a travelling-solvent-floating-zone (TSFZ) technique. For details on the growth technique and the sample characterization, see Ref.\,\cite{Hanna2021}. According to these characterization studies, the crystals grown by the TSFZ method may contain a non-magnetic foreign phase of up to 8 \% volume fraction. In contrast, no such foreign phase was observed for crystals grown by the TSSG method. For both types of crystals a cubic structure with space group P4$_{1}$32 and lattice parameter $a = 12.497 \,\text{\AA}$ for the TSSG crystals and $a = 12.502 \,\text{\AA}$ for the TSFZ crystals was found at room temperature, consistent with earlier reports \cite{Koteswararao2014}.\\
\indent
The labelling of the single crystals (sc) investigated here is identical to that used in Ref.\,\cite{Thurn2021} and consists of a batch number followed by a letter to identify individual samples. The samples sc \#5(b) and sc \#5(c) were grown by using the TSSG method while samples sc \#2 and sc \#1(a) were synthesized via the TSFZ technique.\\
\indent
For measurements of the specific heat, $C(T)$, a homemade calorimeter was used, which is thermally coupled to the mixing chamber of a bottom-loading $^{3}$He-$^{4}$He dilution refrigerator. Measurements were carried out by employing either a thermal relaxation method or a continuous heating method. See \textbf{Appendix} \ref{AppendixA} for details on the data analysis.\\
\indent
The setup for the specific heat measurements was also used for continuous measurements of the magnetic Gr\"uneisen parameter $\Gamma_{B}=1/T\cdot(\partial T/\partial B)_{S}$. For these experiments, the external field was ramped in a certain field window while the temperature of the mixing chamber was slowly varied to match the sample temperature, thereby minimizing the parasitic heat flow, resulting in almost ideal adiabatic conditions. To check for the adiabatic character of these experiments discrete measurements of $\Gamma_{B}$ were performed for selected fields and temperatures. These measurements were conducted by applying a small field step $\Delta B$ while measuring the accompanied temperature change $\Delta T$. See supplemental material from \citep{Thurn2021} and \textbf{Appendix} \ref{AppendixB} for details about the data processing.\\
\indent
Measurements of the coefficient of linear thermal expansion, $\alpha (T)$ = $L^{-1}\partial_{T}\Delta L(T)$, were performed by using two different homemade capacitive dilatometers, depending on the temperature range investigated. Here, relative length changes $\Delta L(T)/L$ were determined via changes of the capacitance where $\Delta L(T) = L(T)-L(T_{0})$ denotes the length change and $T_{0}$ the starting temperature of the experiment. For details of the measurement technique we refer to Refs.\,\cite{Thurn2021, Pott1983}. A bottom-loading $^{3}$He-$^{4}$He dilution refrigerator was employed for investigations of $\alpha (T)$ at low temperatures $0.05 \,\text{K} \leq T \leq 1.7 \,\text{K}$ whereas a conventional $^{4}$He-bath cryostat was used for temperatures $1.5 \,\text{K} \leq T \leq 10 \,\text{K}$. In the dilatometer cells used, the spring leafs exert a force on the crystal, corresponding to an uniaxial pressure along the measuring direction, depending on the sample's dimensions and the starting capacitance used. The applied uniaxial pressure acting along the measuring direction were estimated to $p = (4.9 \pm 0.4)\,\text{MPa}$ for single crystal sc \#5(c), $p = (0.9\pm 0.1)\,\text{MPa}$ for sc \#2, and $p = (0.7\pm 0.1)\,\text{MPa}$ for sc \#1(a).\\
\indent
Magnetostriction experiments were carried out using the same setup as for the thermal expansion measurements. The magnetostriction coefficient $\lambda(B) = L^{-1} \partial_{B}(\Delta L(B))_{T}$ was determined by slowly ramping the field at $T = \text{const.}$ conditions while simultaneously recording the accompanied relative length change $\Delta L(B) = L(B)-L(0)$.\\
\indent
For the experimental investigation of the dielectric permittivity, $\varepsilon'(T)$, two electrodes were established via silver paste on opposite parallel surfaces of the single crystalline samples. The capacitance of this plate capacitor geometry was determined using a LCR meter (Agilent E4980) for different frequencies between $20 \,\text{Hz}$ and $2 \,\text{MHz}$. The dielectric constant was then calculated from the measured capacitance and the sample geometry. Measurements were performed by using a top-loading $^{3}$He-$^{4}$He dilution refrigerator.\\
\indent
The AC magnetic susceptibility, $\chi_{\text{AC}}(T)$, was measured by using a homemade a.c.\,susceptometer adapted to a top-loading $^{3}$He-$^{4}$He dilution refrigerator. Measurements were carried out at a constant frequency of $81 \,\text{Hz}$. To calibrate the setup, magnetization measurements at the highest accessible temperatures of 1.8\,K were compared (by extrapolation) to data taken by using a SQUID magnetometer (Quantum Design MPMS) operating down to 2\,K.
\section{Results}
\subsection{Temperature-dependent experiments}
\begin{figure}[t!]
\includegraphics[width=0.46\textwidth]{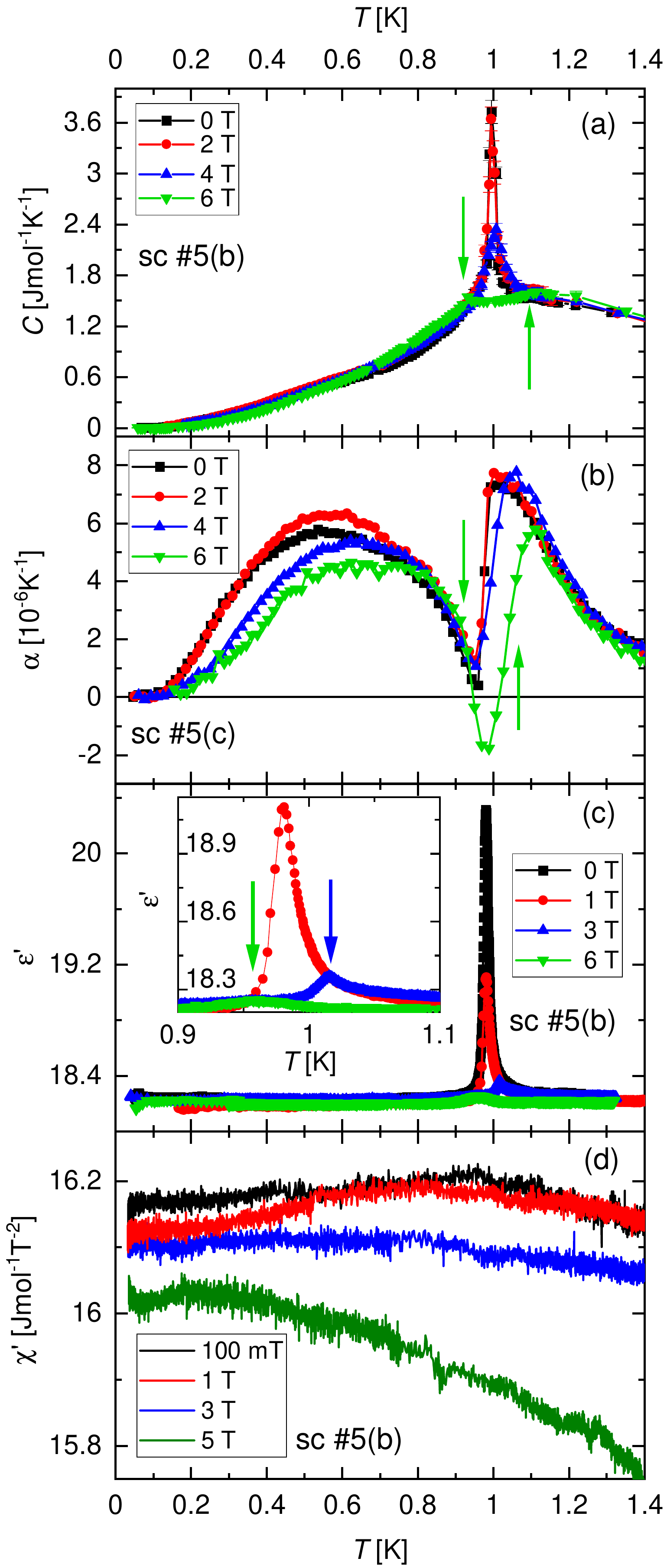}
\caption{Experimental results on single crystalline PbCuTe$_{2}$O$_{6}$ obtained at magnetic fields $B \leq 6\,\text{T}$ applied parallel to [110]. Temperature dependence of (a) specific heat $C(T)$; (b) coefficient of thermal expansion $\alpha(T)$ along [110]; (c) real part of the dielectric permittivity $\varepsilon'(T)$ taken at a constant frequency of $23\,\text{kHz}$ and (d) real part of the AC susceptibility $\chi'(T)$. The inset of (c) represents a zoom-in of the peak anomaly in $\varepsilon'(T)$. Arrows mark specific anomalies for clarification.}
\label{FigLowBThermo}
\end{figure}

\begin{figure}[t!]
\includegraphics[width=0.46\textwidth]{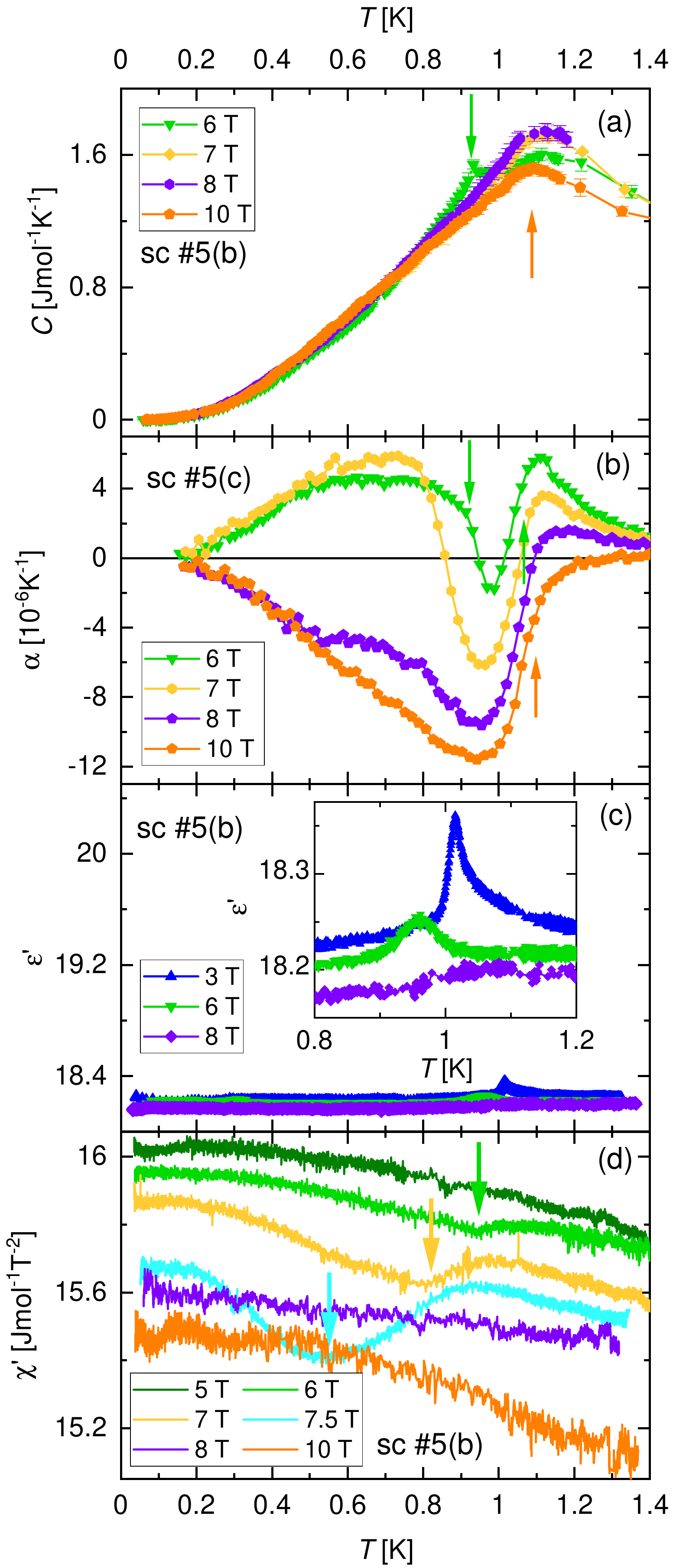}
\caption{Experimental results on single crystalline PbCuTe$_{2}$O$_{6}$ acquired at various magnetic fields $5\,\text{T} \leq B \leq 10\,\text{T}$ applied parallel to [110]. (a) specific heat $C(T)$; (b) coefficient of thermal expansion $\alpha(T)$ along [110]; (c) real part of the dielectric permittivity $\varepsilon'(T)$ taken at a constant frequency of $23\,\text{kHz}$ and (d) real part of the AC susceptibility $\chi'(T)$. A zoom-in on the small peak anomaly in $\varepsilon'(T)$ is displayed in the inset of (c). Arrows indicate the position of selected anomalies.}
\label{FigMidBThermo}
\end{figure}

\begin{figure}[t!]
\centering
\includegraphics[width=0.46\textwidth]{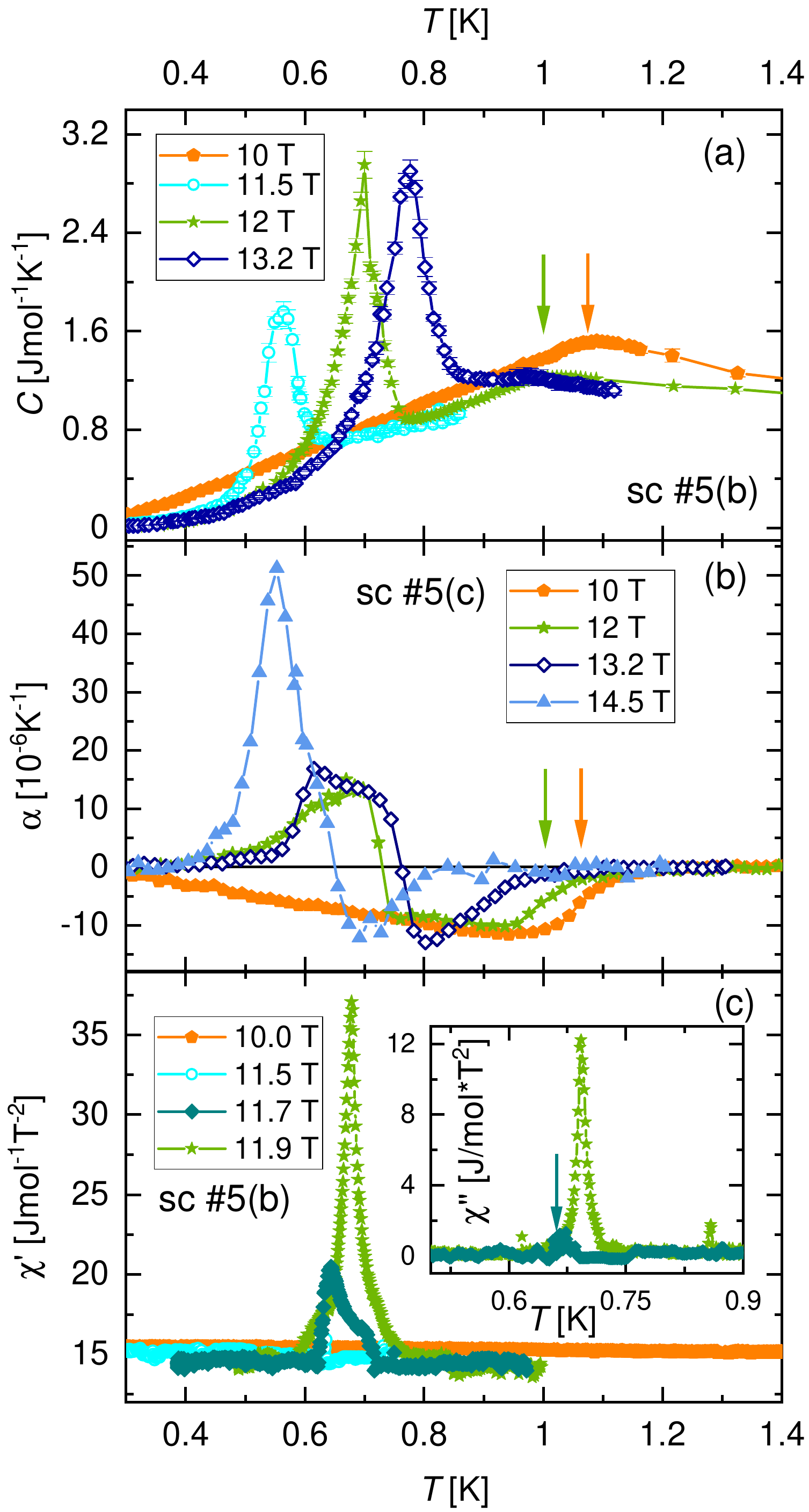}
\caption{Compilation of experimental results on single crystalline PbCuTe$_2$O$_6$ obtained at varying magnetic fields $10\,\text{T} \leq B \leq 14.5\,\text{T}$ applied parallel to the [110] direction. Temperature dependence of (a) specific heat $C(T)$; (b) thermal expansion coefficient $\alpha(T)$ along [110]; (c) real part of the AC susceptibility $\chi'(T)$. The inset of (c) shows the imaginary part $\chi''(T)$. A small peak for $B = 11.7\,\text{T}$ is marked by an arrow.}
\label{FigHighBThermo}
\end{figure}

The full set of data obtained by the various temperature-dependent
measurements performed in this study can be divided into three field regions, including zero-to-moderate fields $B \leq 6\,\text{T}$ [FIG. \ref{FigLowBThermo}], intermediate fields $6\,\text{T} < B \leq 10\,\text{T}$ [FIG. \ref{FigMidBThermo}] and high magnetic fields $B > 10\,\text{T}$ [FIG. \ref{FigHighBThermo}].
\\
\indent
The specific heat data for $B \leq 6\,\text{T}$ are shown in FIG.\,\ref{FigLowBThermo}(a). The $B = 0$ data, already presented in Ref.\,\cite{Thurn2021}, display a sharp lambda-shape anomaly slightly below $1\,\text{K}$, which was attributed to a second-order phase transition to a ferroelectrically-ordered state. The phase-transition anomaly lies on top of a broad maximum which was assigned to the low-energy excitations of the frustrated quantum spin system \cite{Thurn2021}, consistent with the observations made in other frustrated quantum magnets \cite{Okamoto2007, Singh2013, Nakatsuji2005}. For small fields $B \leq 2\,\text{T}$ no significant change can be observed, whereas at higher fields of $4\,\text{T}$ the lambda-shape anomaly somewhat broadens, shrinks in size and slightly shifts to higher temperatures. In these fields $B \leq$ 4\,T there is no discernable effect of $B$ on the broad maximum in $C(T)$. In a field of $6\,\text{T}$, however, a qualitative change becomes obvious as the data reveal indications for two small closely-spaced features around $T_{c1} \approx 0.9\,\text{K}$ and $T_{c2} \approx 1.1\,\text{K}$, cf.\,arrows in FIG.\,\ref{FigLowBThermo}(a), and indications for small changes of the $C(T)$ maximum. In the coefficient of thermal expansion $\alpha(T)$ [FIG. \ref{FigLowBThermo}(b)] a sharp negative anomaly in zero field around $T \approx 1.0 \,\text{K}$ is observed which is superimposed to a broad maximum, cf.\,Ref.\,\cite{Thurn2021}. As revealed for the specific heat, the application of small fields $B \leq 2\,\text{T}$ leaves the phase transition anomaly in $\alpha(T)$ essentially unchanged, whereas a broadening of the negative anomaly and a shift to higher temperatures become evident at $B = 4\,\text{T}$ and $6\,\text{T}$. A comparison of the $6\,\text{T}$ data with the results of $C(T,B = 6\,\text{T})$ in FIG.\,\ref{FigLowBThermo}(a) indicates that the pronounced drop in $\alpha(T)$ [up arrow in FIG.\,\ref{FigLowBThermo}(b)] can be assigned to the transition at $T_{c2}$. In contrast there is only a small feature -- a break in the slope in $\alpha(T)$ -- visible at $T_{c1}$ [down arrow in FIG.\,\ref{FigLowBThermo}(b); see also FIG.\,\ref{FigMidBThermo}(b)]. Figure\,\ref{FigLowBThermo}(c) represents the real part of the dielectric permittivity $\varepsilon'(T)$. The sharp peak in $\varepsilon'(T)$ at $T_{\text{FE}}\approx 1\,\text{K}$, signalling the ferroelectric order \cite{Thurn2021}, is progressively suppressed in size with increasing field, yielding only a small feature at 6\,T [see inset of FIG.\,\ref{FigLowBThermo}(c)]. The peak position, which remains unaffected on increasing the field to $B = 1\,\text{T}$, is slightly shifted to higher temperatures for $B = 3\,\text{T}$ before becoming reduced in height for $B = 6\,\text{T}$ [inset of FIG.\,\ref{FigLowBThermo}(c)]. The real part of the AC susceptibility $\chi'(T)$ is shown in FIG.\,\ref{FigLowBThermo}(d) for fields $B\leq 5\,\text{T}$. The data lack any indication for a phase transition in this temperature- and field range, consistent with the absence of long-range magnetic order. Yet, the data disclose some variations with field, yielding a broad maximum around 0.9\,K for $B = 0.1\,\text{T}$ and $1\,\text{T}$, which becomes suppressed for $B = 3\,\text{T}$. At $B = 5\,\text{T}$ the data reveal a pronounced reduction of $\chi'(T)$ with increasing temperature, indicating a Curie-Weiss-like behavior.\\
\indent
Figure \ref{FigMidBThermo}(a) illustrates the specific heat data for intermediate fields of $6\,\text{T} \leq B \leq 10\,\text{T}$. For $B = 8\,\text{T}$ and $10\,\text{T}$ no sharp phase transition anomaly can be resolved anymore. Instead $C(T)$ shows a rounded maximum, the height of which first increases slightly on going from $B = 6\,\text{T}$ to $8\,\text{T}$. At $B = 10\,\text{T}$ it becomes reduced in size and slightly shifted to lower temperature. More strongly pronounced anomalies in this intermediate field range were found in the coefficient of thermal expansion [FIG.\,\ref{FigMidBThermo}(b)].
Upon increasing the field from $B = 6\,\text{T}$ to $7\,\text{T}$ the negative anomaly in  $\alpha(T)$ around $T = 1\,\text{K}$ increases in size giving rise to a rounded minimum, whereas the positive contribution at lower temperatures is almost unaffected. On further increasing the field to $8\,\text{T}$ and $10\,\text{T}$, however, this positive contribution becomes removed leaving behind a large negative anomaly in $\alpha(T)$, characterized by a slightly rounded step-like change on its high-temperature flank around $T \sim 1.1\,\text{K}$, reminiscent of a broadened mean-field-type phase transition. As becomes evident from FIG.\,\ref{FigMidBThermo}(c), this pronounced feature in $\alpha(T)$ at $B=8\,\text{T}$ has no corresponding signature in the dielectric permittivity $\varepsilon'(T)$, reflecting a complete suppression of the peak anomaly, revealed at lower fields [inset Fig.\,\ref{FigMidBThermo}(c)]. No further anomaly could be observed in $\varepsilon'$ in the temperature window investigated for fields $ B \leq 10\,\text{T}$. In contrast, some indications for anomalous behavior are revealed in this intermediate field range in the magnetic susceptibility $\chi'(T)$ [FIG.\,\ref{FigMidBThermo}(d)]. This anomaly evolves from a small dip slightly above $0.9\,\text{K}$ in the data taken at $B=6\,\text{T}$. Upon increasing the field to $7\,\text{T}$ and $7.5\,\text{T}$ the anomaly grows in size, broadens and becomes shifted to lower temperatures. On further increasing the field to $8\,\text{T}$ and $10\,\text{T}$, this anomaly disappears and no other anomalous feature can be resolved in the temperature window investigated. We note that in the field range 5\,T $\leq B \leq$ 10\,T no corresponding anomaly could be observed in the imaginary part $\chi''(T)$ (not shown) within the experimental resolution.
\\
\indent
Another drastic change in the material's thermodynamic and magnetic properties becomes apparent on further increasing the field to above $B=10\,\text{T}$ [FIG.\,\ref{FigHighBThermo}]. While the rounded maximum of $C(T)$ revealed at $10\,\text{T}$ continues to further shrink in size and to shift to lower temperatures with increasing the field to $12\,\text{T}$, a pronounced phase transition anomaly becomes visible for fields $B\geq 11.5\,\text{T}$ [FIG.\,\ref{FigHighBThermo}(a)]. This anomaly continuously shifts to higher temperatures upon increasing the field from 11.5 to 12 and $13.2\,\text{T}$ while growing in size. A large corresponding phase transition anomaly is also revealed in the coefficient of thermal expansion $\alpha(T)$, manifesting itself in a pronounced positive peak [FIG. \ref{FigHighBThermo}(b)]. This phase transition anomaly sits on top of a negative contribution, resembling the anomaly revealed at $B=10\,\text{T}$. For fields around $B=13\,\text{T}$ it develops a substructure, indicating a split into two features, i.e., a rather sharp peak on the low-temperature side and a kink followed by a drop on the high-temperature side. On further increasing the field to $14.5\,\text{T}$, a single strongly pronounced sharp anomaly is observed. Figure\,\ref{FigHighBThermo}(c) illustrates the magnetic susceptibility $\chi'(T)$ in this high-field region. The data reveal a sharp peak anomaly which starts to become visible at $B=11.5\,\text{T}$, and grows significantly in size on increasing the field to $11.7\,\text{T}$ and $11.9\,\text{T}$. This growth is accompanied by a shift of the peak position to higher temperatures. As demonstrated in the inset of FIG.\,\ref{FigHighBThermo}(c), a corresponding anomaly is also revealed in the imaginary part of the magnetic susceptibility $\chi''(T)$.
\subsection{Field-dependent experiments}
\begin{figure}[b!]
\centering
\includegraphics[width=0.47\textwidth]{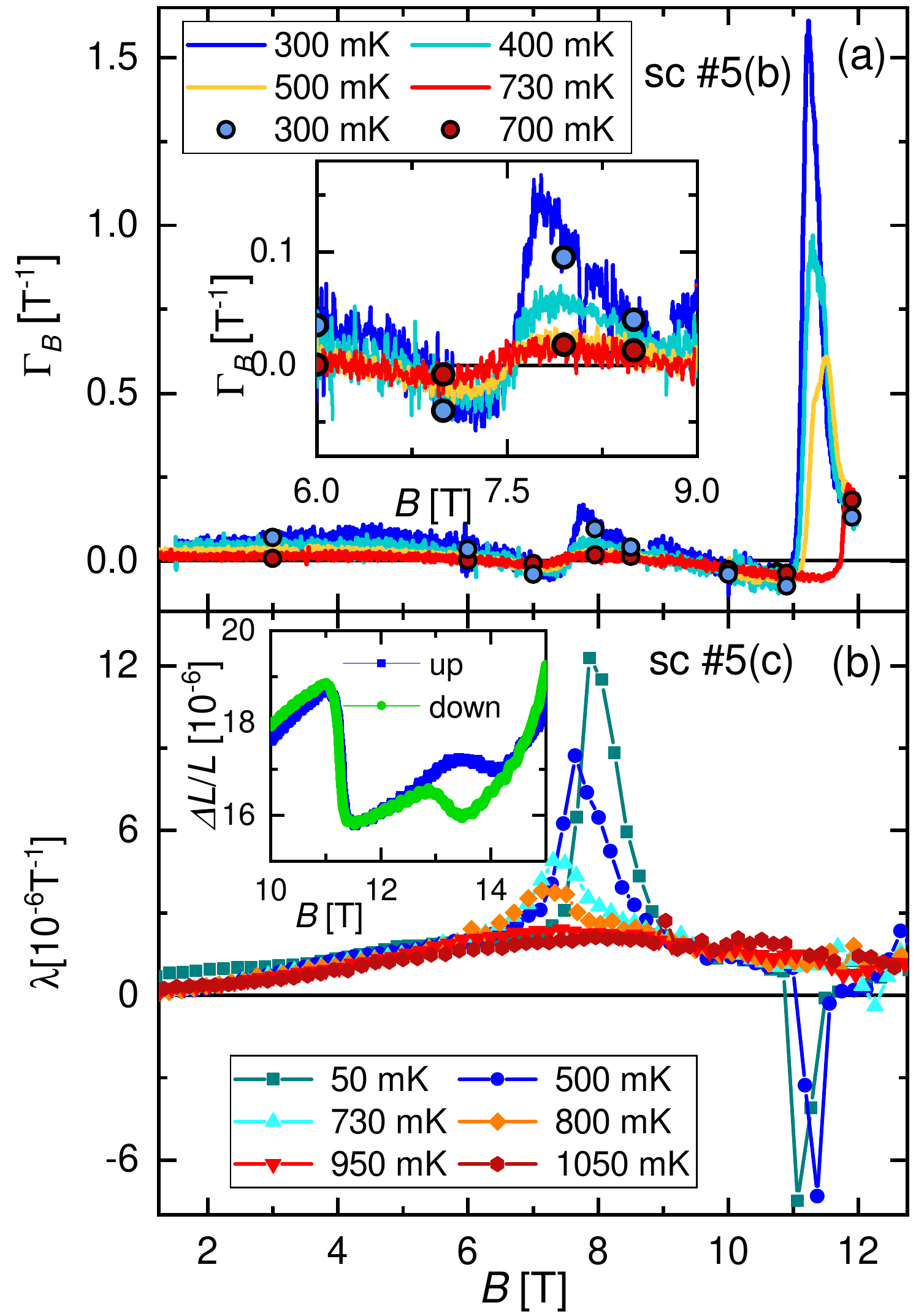}
\caption{Compilation of experimental results on single crystalline PbCuTe$_{2}$O$_{6}$ obtained by sweeping the magnetic field $B$ $\vert\vert$ [110] in the range $1\,\text{T} \leq B \leq 13\,\text{T}$. (a) Magnetic Gr\"uneisen parameter $\Gamma_{B}(B)$ for field-up measurements as a function of field $B$. As explained in the text the stated temperature represents an average value at which individual field sweeps were performed. In addition, discrete data points of $\Gamma_{B}(B)$ are shown for selected temperatures and fields. The corresponding error bars are within symbol size. (b) Magnetostriction coefficient $\lambda(B)$ along [110] vs $B$ for field-up measurements at different temperatures. The inset shows a section of the corresponding relative length change $\Delta L(B)/L$ for a representative measurement at $T = 0.5\,\text{K}$.}
\label{FieldExp}
\end{figure}
The temperature-dependent studies [FIGs.\,\ref{FigLowBThermo}-\ref{FigHighBThermo}] were complemented by experiments performed under variation of magnetic field. In FIG.\,\ref{FieldExp}(a) we show the results obtained from continuous measurements of the magnetic Gr\"uneisen parameter $\Gamma_{B}(B)$ taken at varying temperatures $T \approx$ const. For these experiments $\Gamma_{B}(B)$ is derived from continuous sweeps over narrow field intervals. Since this technique implies some deviations from ideal $T = \text{const.}$ conditions, the stated temperatures represent averaged values (see \textbf{Appendix} \ref{AppendixB} for details). In order to check for the adiabatic character of these experiments, additional data points, acquired via discrete measurements, are also plotted for selected temperatures and magnetic fields in FIG.\,\ref{FieldExp}(a). The good agreement of both data sets indicates sufficiently good adiabatic conditions in the continuous experiments. The data for $\Gamma_{B}(B)$ taken at a temperature of $T \approx 0.73\,\text{K}$ show a flat, almost $B$-independent behavior for $B < 6 \,\text{T}$. Upon further increasing the magnetic field an S-shaped anomaly, i.e., a minimum-maximum structure for $6\,\text{T} \leq B \leq 8\,\text{T}$ emerges, which is accompanied by a change of sign around $B_{c1} = 7.4\,\text{T}$. The peaks become more pronounced upon decreasing the temperature down to 0.3\,K while the sign change slightly shifts to higher fields, [see inset to  FIG.\,\ref{FieldExp}(a)]. A small hysteresis between field-up and field-down measurements (not shown) can be observed for this feature. Since no indications for latent heat could be revealed in any of the temperature-dependent experiments in the corresponding $B$-$T$ parameter range, we refrain from attributing the hysteresis to a first-order phase transition. Rather, we assign this observation to non-equilibrium, i.e., thermalization effects. Further experiments to be performed by systematically varying the field rates are necessary to clarify the character of the transition. Above $B = 11 \,\text{T}$ yet another change of sign is observed followed by a sharp jump-like increase in $\Gamma_{B}(B)$ at $B_{c2}=11.7\,\text{T}$ for $T \approx 0.73\,\text{K}$. Upon decreasing temperature, the position of this sign change is reduced, reaching $11.2 \,\text{T}$ at $T \approx 0.3\,\text{K}$, while the size of the peak grows significantly. Note that a clear identification of potential hysteretic behavior in this field range is hampered due to extrinsic factors as discussed in \textbf{Appendix} \ref{AppendixB}. \\
\indent
In FIG.\,\ref{FieldExp}(b) we show results of magnetostriction measurements $\lambda(B,T\approx \text{const.})$ performed along the [110] direction. In the main panel only data taken upon increasing the field are shown for clarity. Whereas a smooth variation of $\lambda(B)$, yielding a broad maximum around 8\,T, is revealed for the highest temperatures of the experiment of $T \approx 0.95\,\text{K}$ and $1.05\,\text{K}$, peak-like anomalies become visible at lower temperatures $T \leq 0.8\,\text{K}$. These include a maximum between $B=6\,\text{T}$ and $B=8\,\text{T}$, depending on the temperature, and a sharp negative peak slightly above $B=11\,\text{T}$. Both features grow in size for decreasing temperatures. This growth is accompanied by a shift to higher fields for the maximum around $7\,\text{T}$ and a small decrease in the position of the high-field anomaly. The data taken upon decreasing the field reveal a small hysteresis for the broad maximum around $B=7\,\text{T}$. In contrast no clear indications for such a hysteresis could be observed for the high-field anomaly at $B=11\,\text{T}$ [see inset of FIG.\,\ref{FieldExp}(b) for the raw data at $0.5\,\text{K}$]. A closer look on the raw data showing the relative length change $\Delta L(B)/L$ in the inset of FIG.\,\ref{FieldExp}(b) discloses yet another feature, besides the main step-like change in $\Delta L(B)/L$ around $11\,\text{T}$. At around $13.5\,\text{T}$ another slightly broadened step-like anomaly is observed which shows a clear hysteresis for field-up and field$\text{-}$down measurements.
\begin{figure}[b!]
\centering
\includegraphics[width=0.48\textwidth]{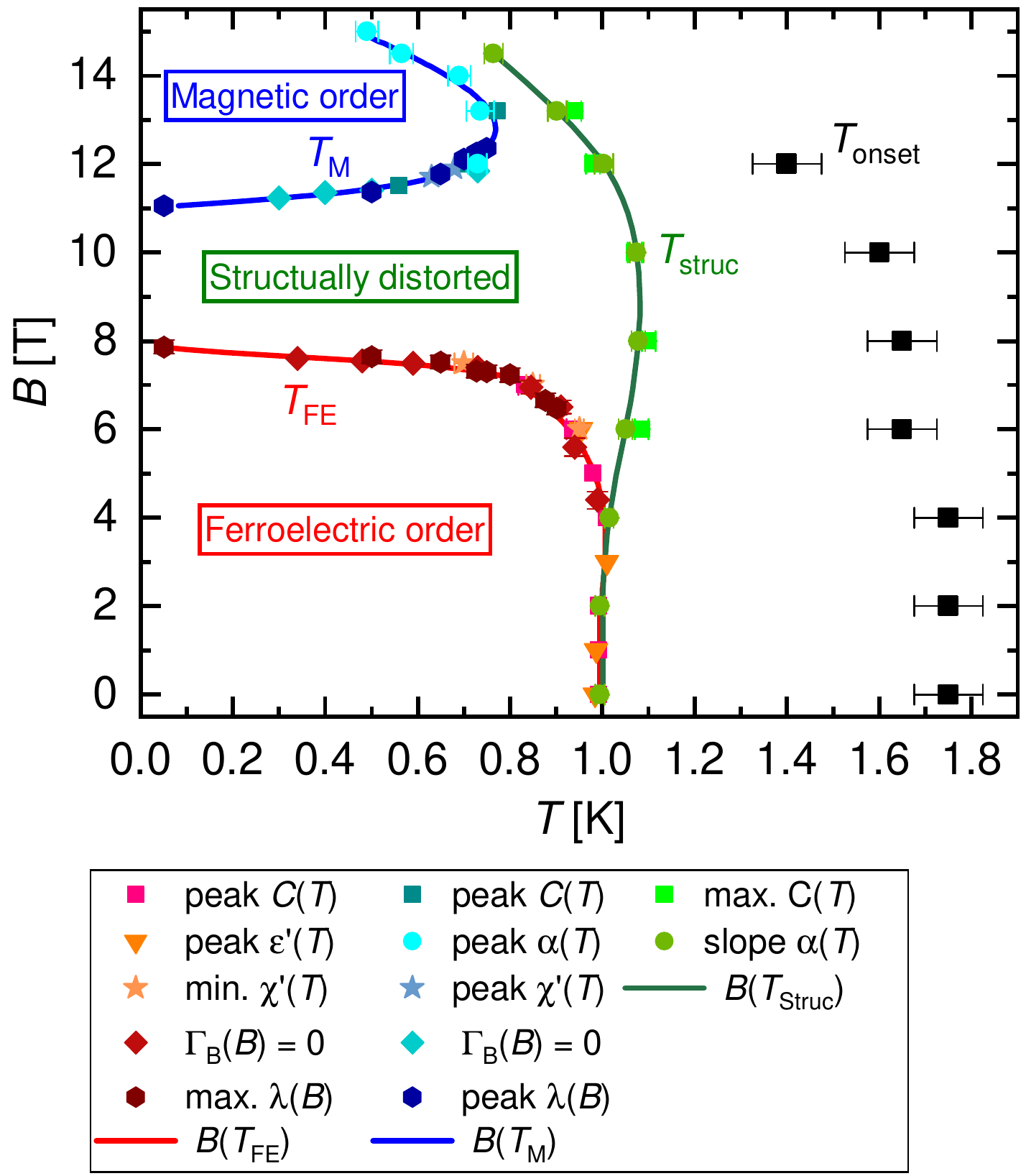}
\caption{$B - T$ phase diagram of single crystalline PbCuTe$_{2}$O$_{6}$ for $B$ $\vert\vert$ [110]. The boundaries for the ferroelectrically-ordered phase $T_{\text{FE}}(B)$, the
structurally-distorted state $T_{struc}(B)$ and magnetically long-ranged-ordered phase $T_{\text{M}}(B)$ are derived from the field- and temperature positions of the different anomalies in the various quantities as indicated in FIGs. 1-4. The precursor for the lattice distortion, setting in at $T \leq T_{\text{onset}}(B)$, is derived from the point below which the $\alpha(T)$ curves for different orientations start to differ, cf.\,FIG.\,\ref{FigDistortion} for data at $B$ = 0. Except for the $B$-induced magnetically ordered state above $11\,\text{T}$, the data revealed here are consistent with a spin liquid state, which has been proposed based on experiments at $B=0$ \cite{Chillal2020}. Representative errors are estimated from the width of the corresponding anomalies or of the width of a possible hysteresis. If not indicated otherwise, the characteristic errors are within symbol size. For further details refer to the text.}
\label{phaseDiag}
\end{figure}
\\
\section{Discussion}
\subsection{Tracing the various phase transition lines}
The various anomalies revealed in the thermodynamic, dielectric and magnetic properties by means of temperature- and field-dependent investigations have been used to construct a $B - T$ phase diagram [FIG.\,\ref{phaseDiag}]. Three different phase transition lines can be distinguished in the field- and temperature range covered here. See the discussion in Sec.\,B below for the assignment of the various phases.\\
\indent
These include (1) the phase transition into ferroelectric (FE) order for $T \leq T_{\text{FE}}$. The corresponding phase transition line $T_{\text{FE}}(B)$ is derived from the peak position of the $\lambda$-shape anomalies in $C(T)$ and $\varepsilon'(T)$ for fields $B \leq 7\,\text{T}$. For the determination of the phase boundary at fields 6\,T $\leq B \leq$ 8\,T, where $T_{\text{FE}}(B)$ strongly changes with $B$, we also include the positions of phase transition anomalies revealed in field-dependent measurements, i.e., the sign change in $\Gamma_{B}(B)$ and the position of the maximum in $\lambda(B)$.
\\
\indent
(2) The phase transition into the FE state around $1\,\text{K}$ is accompanied by a sharp negative anomaly in the coefficient of thermal expansion $\alpha(T)$, indicating a structurally distorted low-temperature state. By identifying  the point of maximum slope in $\alpha$ with the phase transition temperature $T_{\text{struc}}$, we find coinciding $T_{\text{struc}}(B)$ and $T_{\text{FE}}(B)$ at low fields $B \leq 4\,T$. At higher fields $B > 4\,\text{T}$, however, $T_{\text{struc}}(B)$ clearly splits off [cf.\,the features at $T_{c1}$ and $T_{c2} = T_{\text{struc}}$ in $C(T, B$ = const.) at $B=6\,\text{T}$ in FIG.\,\ref{FigLowBThermo}(a)] and becomes even slightly enhanced upon increasing the field to about 10\,T. The phase diagram in FIG.\,\ref{phaseDiag} suggests the possibility of a bi-critical point around 1 K and 4 T. However, since the phase transitions associated with $T_{\text{FE}}$ and $T_{\text{struc}}$ for $B > 4\,\text{T}$ are of second order, such a scenario would imply a single first-order transition line for $B < 4\,\text{T}$. This is not supported by our experiments, where indications were found neither for latent heat nor for discontinuous length changes. For fields in excess of 10\,T, $T_{\text{struc}}(B)$ becomes strongly reduced suggesting a full suppression for fields in the range of about 18\,T. Note that for fields $B > 6\,\text{T}$ the thermal expansion anomaly at $T_{\text{struc}}(B)$ coincides with the position of the rounded peak in $C(T)$.\\
\indent
(3) At high fields of $B \geq 11\,\text{T}$, another phase transition anomaly is observed, which is assigned to a transition into long-range magnetic order, see below. The corresponding phase boundary $T_{\text{M}}(B)$ is derived by using the peak position of the pronounced phase transition anomalies in $C(T)$, $\chi'(T)$ and $\alpha(T)$. For fields around 11\,T where the phase boundary is flat, the phase transition line $T_{\text{M}}(B)$ also includes the position of the sign change in $\Gamma_{B}(B)$ as well as the sharp negative peak in $\lambda(B)$ for $B \geq 11\,\text{T}$.
\subsection{Assignment of the various phases}

\textbf{Ferroelectric order} The assignment of the state below $T_{\text{FE}}(B)$ to ferroelectric order is based on detailed investigations performed at $B = 0$ in Ref.\,\cite{Thurn2021} of the dielectric constant $\varepsilon'(T)$ and electric polarization $P(T)$. There it was shown via measurements of the pyrocurrent that the peak in $\varepsilon'(T)$ coincides with the formation of a macroscopic electric polarization $P$, which can be switched depending on the applied electrical field \cite{Wieder1955, Thurn2021}. Furthermore, from frequency-dependent measurements the ferroelectric transition was found to be of order-disorder type \cite{Thurn2021}, with small sample-to-sample variations in the height of the peak in $\varepsilon'$. By following this peak in $\varepsilon'(T, B=\text{const.})$, we find that the ferroelectric state is also stable in finite fields albeit with diminishing signature in $\varepsilon'(T)$ and $\varepsilon''(T)$ (not shown) on increasing the magnetic field. No indications for irreversible effects depending on the field history were found. The phase diagram in FIG.\,\ref{phaseDiag} suggests a full suppression of $T_{\text{FE}}(B)$ to $T=0$, i.e., a field-induced ferroelectric quantum critical point, for fields around $B_{c1} = 7.9 \,\text{T}$. The notion of a quantum critical point at $B_{c1}$ implies a second-order phase transition at $T_{\text{FE}}$. In fact, the results obtained in $B = 0$ in Ref.\,\cite{Thurn2021} and those in finite fields presented here all are consistent with a second-order phase transition at $T_{\text{FE}}$. This notion also includes the weak asymmetry revealed for $\Gamma_{B}$ around $B_{c1}$ [inset of FIG.\,\ref{FieldExp}(a)], as observed also for other quantum critical systems \cite{Wolf2011}. To fully characterize the quantum critical point at $B_{c1}$ further experiments are necessary which also enable a scaling analysis of $\Gamma_{B}$ to be performed. The emergence of a quantum critical point is generally associated with the suppression of a characteristic energy scale at the QCP \cite{Zhu2003}. Since the ferroelectric transition is accompanied by significant distortions of the crystal lattice, as observed by $\lambda(B)$ [see FIG.\,\ref{FieldExp}(b)], this energy scale is likely to be assigned to a combination of dielectric- and lattice degrees of freedom. In the high-field region of the ferroelectric state, i.e., for fields 5\,T $\leq B \leq$ 7.9\,T, we find that crossing the phase boundary $T_{\text{FE}}(B)$ is accompanied by significant changes in the crystal lattice [compare FIG.\,\ref{FigMidBThermo}(b) and \ref{FieldExp}(b)]. In light of the material's considerably strong magneto-elastic coupling (see discussion below), we expect that these changes are responsible for the signatures revealed in the magnetic susceptibility in this field range [FIG. \ref{FigMidBThermo}(d)]. This notion is consistent with the absence of any feature in $\chi''$ on crossing $T_{\text{FE}}(B)$, as opposed to the appearance of peak anomalies in $\chi''$ on crossing $T_{\text{M}}(B)$ for $B\geq 11\,\text{T}$, [see FIG. \ref{FigHighBThermo}(c)].
\\

\indent
\textbf{Field-induced magnetic order} The assignment of the high-field phase for $B \geq$ 11\,T and $T \leq T_{\text{M}}$ to long-range magnetic order is based on the pronounced peaks revealed in the magnetic susceptibility $\chi'(T, B$ = const.). In addition, the large anomalies observed in the specific heat, corresponding to the release of a significant amount of entropy of $ \Delta S \approx 0.07\cdot\text{Rln(2)}$ within a narrow temperature interval of $\sim 0.15\,\text{K}$ around $T_M$, are consistent with the ordering of magnetic degrees of freedom. The fact that DFT calculations show exclusively antiferromagnetic interactions \cite{Chillal2020} suggests the magnetic order to be of antiferromagnetic nature. In contrast to the ferroelectric transition, which is of second order, the $B$-induced magnetic transition shows indications of a first-order transition. This refers to the discontinuous change revealed in the magnetostriction [inset of FIG. 4(b)] and the pronounced asymmetry of the magnetic Grüneisen parameter at the critical field, see Ref.\,\cite{Tokiwa2013}. As a consequence, we refer to the point where $T_{\text{M}}$ is suppressed to zero as a quantum phase transition. In this context we assign the anomaly in $\Delta L(B)/L$ around $13.5\,\text{T}$ at $T=0.5\,\text{K}$ [inset of FIG.\,\ref{FieldExp}(b)] together with the substructure in $\alpha(T)$ around $0.6\,\text{K}$ for $B=13.2\,\text{T}$ [FIG.\,\ref{FigHighBThermo}(b)], both of which fall into the region of the magnetically-ordered phase, to spin reorientation transitions. The phase diagram in FIG.\,\ref{phaseDiag} indicates a field-induced magnetic quantum phase transition at $B_{c2}$ = 11\,T.
\begin{figure}[t!]
\centering
\includegraphics[width=0.47\textwidth]{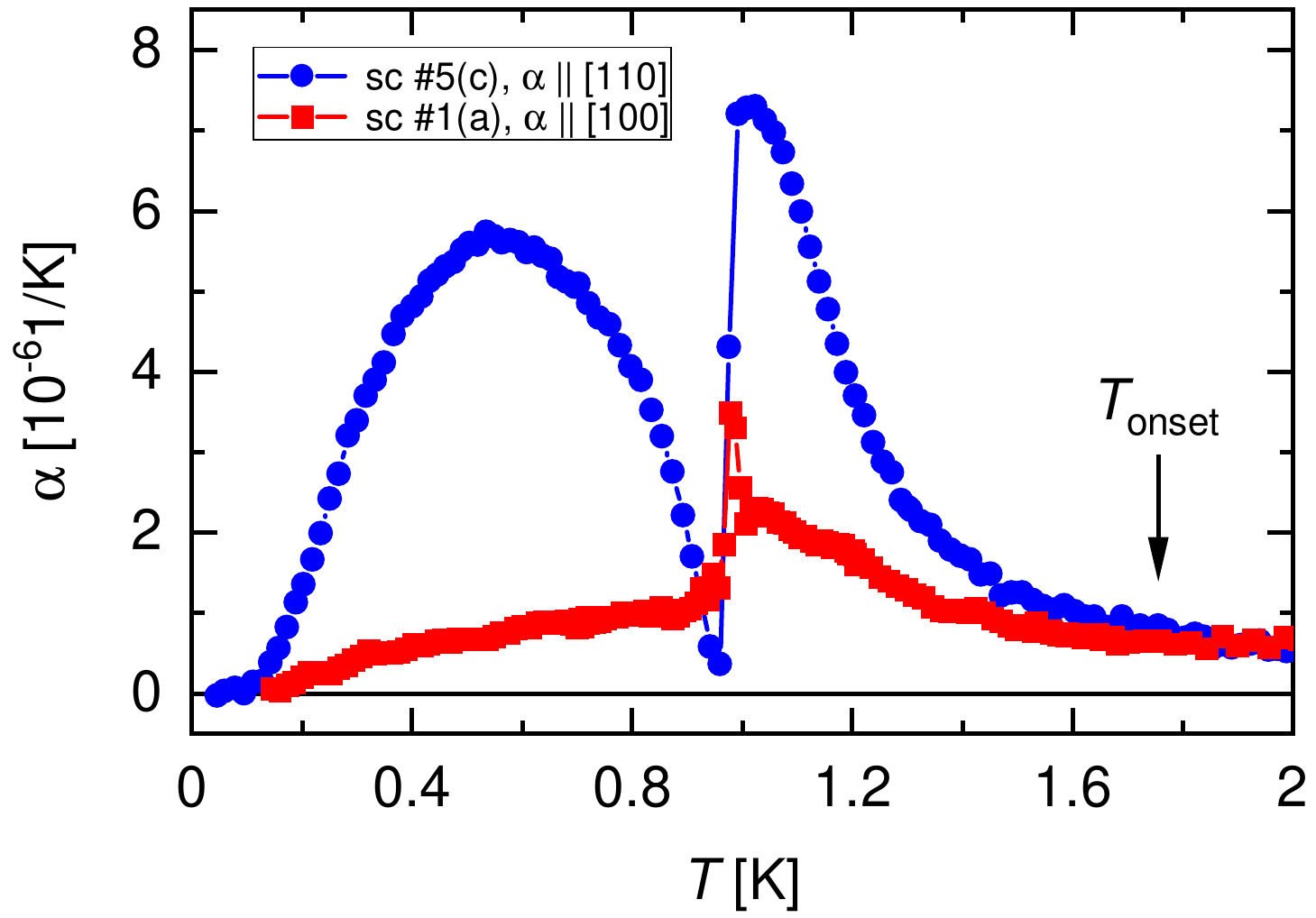}
\caption{Temperature dependence of the thermal expansion coefficient of PbCuTe$_{2}$O$_{6}$ in zero field measured on single crystal sc \#5(c) along [110] and on sc \#1(a) along [100]. The arrow marks the temperature $T_{\text{onset}}$ below which both data sets start to split signalizing the onset of anisotropic lattice distortions.}
\label{FigDistortion}
\end{figure}
\\
\indent

\textbf{Structurally-distorted phase}
The regions of ferroelectric order and magnetic order are bounded by the transition into the structurally-distorted non-cubic phase. This assignment is based on observations made in thermal expansion measurements performed on single-crystalline samples along two different directions, identified as [100] and [110] at room temperature, see FIG.\,\ref{FigDistortion} and Ref.\,\cite{Thurn2021}. Whereas the two data sets coincide at higher temperatures down to about 1.75\,K, consistent with cubic symmetry, it is found that upon further cooling to temperatures below $T_{\text{onset}}$ = 1.75\,K the system develops an anisotropic lattice distortion, reflecting deviations from a cubic symmetry. As discussed in detail in Ref.\,\cite{Thurn2021}, this deviation from cubic symmetry implies the formation of structural domains which can be influenced by the uniaxial pressure exerted by the dilatometer cell. We interpret the growth in $\alpha(T)$ below about 1.75\,K and the accompanying anisotropic lattice distortion as a precursor of the structural transition at $T_{\text{struc}}$ around 1\,K. This interpretation is corroborated by following the onset of anisotropic behavior at $T \leq T_{\text{onset}}(B)$ in finite fields, c.f., the black squares in FIG.\,\ref{phaseDiag}. The so-derived $T_{\text{onset}}(B)$ appears to preempt the evolution of $T_{\text{struc}}(B)$ in the $B$-$T$ range investigated, in support of the above interpretation. The significant reduction of both $T_{\text{onset}}(B)$ and $T_{\text{struc}}(B)$ for a field of 12\,T signals the presence of a significant magneto-elastic coupling in this material, see Sec.\,C below. \\

\textbf{Quantum-disordered, potentially quantum spin liquid state}
According to the magnetic measurements performed in the present study and in Ref.\,\cite{Thurn2021} single crystalline PbCuTe$_2$O$_6$ lacks any indication for long-range magnetic order at fields $B <$ 11\,T and temperatures down to $T=0.04\,\text{K}$. Detailed investigations of the magnetization $M(T)$ for these single crystals in Ref.\,\cite{Thurn2021} showed a mild reduction with decreasing temperatures below about 0.8\,K, see also the susceptibility data shown in FIG.\,\ref{FigMidBThermo}(d). This contrasts with the slight increase in $M(T)$ upon cooling below 1\,K revealed for the pressed-powder samples \cite{Khuntia2016}. This slightly modified magnetic response for the single crystals, which is still consistent with a gapless QSL, has been assigned to the lattice distortions accompanying the ferroelectric order. These lattice distortions may in turn alter the magnetic interactions between the Cu$^{2+}$ ($S$ = 1/2) ions. See \textbf{Appendix} \ref{AppendixC} for more details on this aspect. Interestingly, by studying the magnetic Gr\"{u}neisen parameter as a function of temperature, indications for quantum critical behavior were revealed for the single crystals \cite{Thurn2021} whereas non-critical behavior characterizes the pressed-powder material. Based on measurements in small magnetic fields of $B\sim 100\,\text{mT}$ it was concluded that single crystalline PbCuTe$_2$O$_6$ for $T \rightarrow 0$ approaches a quantum critical point which is sensitive to magnetic fields \cite{Thurn2021}.

%Similarly, clear indications for the divergence of $\Gamma_{B}(T)$ for $T \rightarrow 0$ were observed (wenn wir das so sagen, erwartet man, dass wir die Daten auch zeigen; wollen wir das?) around $B = 7.9\,\text{T}$ and $B = 11\,\text{T}$. Although these signatures would likewise signal a quantum critical point at the end of the phase separation line for the ferroelectrically and for the magnetically ordered state, a precise quantitative analysis of $\Gamma_{B}(T)$ at lowest temperatures $T\leq 0.2\,\text{K}$ is hampered due to the uncertainty in the correction of the experimental data for the contribution of the nuclear, magnetic moments (see \textbf{Appendix} \ref{AppendixB}).

\subsection{Effect of magnetic field on ferroelectric order}
\label{CouplingBdielec}
The observation of ferroelectric order in PbCuTe$_{2}$O$_{6}$ which can be suppressed by sufficiently strong magnetic fields raises the question of how the magnetic field couples to the dielectric degrees of freedom in this system. To shed light on the nature of the coupling at work here, it is instructive to study the material's elastic properties at temperatures $T \geq 1.7\,\text{K}$ and how these elastic properties respond to a magnetic field.
%\indent
\begin{figure}[t!]
\centering
\includegraphics[width=0.47\textwidth]{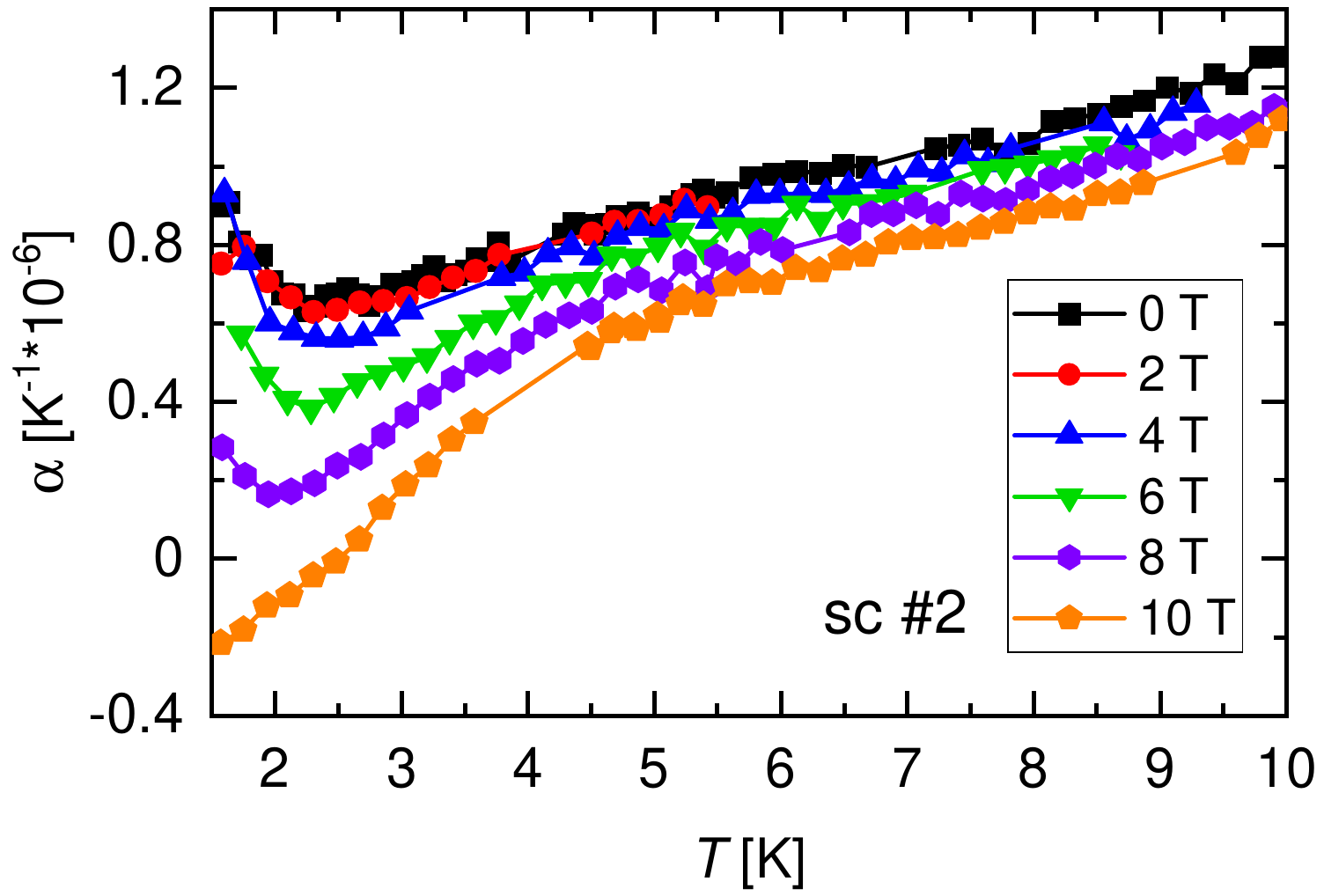}
\caption{Temperature dependence of the thermal expansion coefficient of single crystalline PbCuTe$_{2}$O$_{6}$ along [110] in varying fields $B$ $\vert\vert$ [110].}
\label{FigExplHighB}
\end{figure}
\\
\indent
To this end we show in FIG.\,\ref{FigExplHighB} data for the coefficient of linear thermal expansion for temperatures 1.5\,K $\leq T \leq$ 10\,K taken at varying fields $B \leq$ 10\,T. The $\alpha(T)$ data along [110] reveal a considerable field dependence for $B >$ 4\,T which becomes particularly pronounced for temperatures $T \leq$ 5\,K. The data indicate that at these temperatures, where the material is still in its non-distorted cubic phase at $T > T_{\text{onset}}$, the application of a sufficiently strong magnetic field gives rise to significant changes in the material's lattice parameter. Pronounced magneto-elastic effects in the paramagnetic regime of Cu$^{2+}$-containing materials are known, see, e.g., Refs.\,\cite{Cong2014,Liu2014}, reflecting a considerably strong spin-orbit interaction of the Cu ions in these systems. As a result, cooling the system in the presence of a magnetic field may give rise to some lattice strain. According to the data in FIG.\,\ref{FigExplHighB} this effect is minor at fields $B \leq$ 4\,T, consistent with the almost vertical slope of $T_{\text{FE}}(B)$ in FIG.\,\ref{phaseDiag}. However, on increasing the field to 6\,T and above, the progressive suppression of $\alpha(T)$ at these fields indicates significant magneto-elastic effects. At around the same fields where these magnetostrictive effects arise, the ferroelectric transition at $T_{\text{FE}}(B)$ and the structural transition at $T_{\text{struc}}(B)$ start to separate [cf.\,FIG.\,\ref{phaseDiag}], suggesting that the $B$-induced lattice distortions act differently on $T_{\text{struc}}(B)$ and $T_{\text{FE}}(B)$: whereas fields in the range 4\,T $\leq B \leq$ 8\,T appear to be advantageous for the structural transition, they are detrimental for the ferroelectric transition. In the supplemental information of Ref.\,\cite{Thurn2021} it was suggested that out of the two non-equivalent Pb ions (Pb1 and Pb2) (see \textbf{Appendix} \ref{AppendixC} for details) characterizing this structure, the Pb2 ions are the ones which are responsible for the ferroelectric ordering. It was argued that atomic displacements which lower the symmetry of the oxygen coordination of these Pb2 ions are key for the formation of lone pairs and thus ferroelectric order. As shown in detail in \textbf{Appendix} \ref{AppendixC} the oxygen O1 atoms represent the link between the stereochemically active Pb2 ions and the magnetic Cu$^{2+}$ ions. We therefore speculate that the $B$-induced lattice distortions are such that they counteract the lone pair formation.\\

\section{Conclusions} %\textendash\ ............................\\

In conclusion, the application of a variety of thermodynamic, magnetic, and dielectric probes to single crystalline PbCuTe$_2$O$_6$ in magnetic fields $B \leq$ 14.5\,T, reveals a rich $B$-$T$ phase diagram at low temperatures $T \leq$ 1.8\,K. The phases observed include ferroelectric order at zero to moderately strong magnetic fields $B \leq$ 8\,T and a field-induced magnetically-ordered state at high fields $B \geq$ 11\,T. These phases are preceded by or coincide with a structural transition from a high-temperature cubic structure to a distorted non-cubic low-temperature state. Our study provide evidence for a ferroelectric quantum critical point at $B_{c1} = 7.9\,\text{T}$, where the second-order phase transition line associated with ferroelectric order is suppressed to zero. In addition, we find indications for a magnetic quantum phase transition at $B_{c2} = 11\,\text{T}$. Detailed structure investigations in zero and finite fields are needed to identify the symmetry of the low-temperature phase in PbCuTe$_2$O$_6$ and to elucidate the accompanied changes in the atomic positions. Of particular interest are changes in the local environment of the material's polar building blocks, i.e., the Te$^{4+}$ ions in the oxotellurate tetrahedra and the two inequivalent Pb$^{2+}$ ions, to clarify their role in the formation of the electric dipoles and to understand the mechanism by which the magnetic field couples to the dielectric- and magnetic degrees of freedom in this material. In lacking such detailed microscopic structural information, indications for the nature of the coupling between the magnetic field and the dielectric degrees of freedom at work here were found in thermal expansion measurements, yielding pronounced $B$-induced lattice distortions at temperatures preceding the symmetry-broken low-temperature phases. These magnetostrictive effects were attributed to the spin-orbit interaction of the Cu$^{2+}$ ions and the effect these local lattice distortions may have on the environment of the Pb2 ions in the material. An interesting consequence of the lattice distortions being coupled to both the magnetic field and the dielectric degrees of freedom would be a finite magneto-electric (ME) effect. This exciting possibility of a strain-mediated ME in PbCuTe$_{2}$O$_{6}$ should be explored by further experiments.

%----------------------%
%  8. Acknowledgement  %
%----------------------%
\section*{Acknowledgments}
Work done at Goethe-University Frankfurt was supported by the German Science Foundation (DFG) through project A06 of SFB/TR 288 (ID 422213477). B.L. acknowledges the support of Deutsche Forschungsgemeinschaft (DFG) through project B06 of SFB 1143: Correlated  Magnetism: From Frustration To Topology (ID 247310070). The crystal growth took place at the Core Lab Quantum Materials, Helmholtz Zentrum Berlin f\"ur Materialien und Energie, Germany.

%\newpage

\appendix

\begin{appendix}

\section{Details of the data evaluation of $C(T)$}
\label{AppendixA}
The acquired specific heat data consist in principle of three different contributions resulting from the crystal lattice, the magnetic moments of the nuclei with non-zero spin and the magnetic moments of the $3d$ electrons of Cu$^{2+}$ ions. The contribution from the crystal lattice is according to Ref.\,\cite{Koteswararao2014} smaller than 1\% of the total specific heat for temperatures $T \leq 1.6\,\text{K}$ and can thus be neglected. The contributions from the nuclear magnetic moments can be calculated as Schottky anomalies due to the Zeeman interaction of the $^{207}$Pb, $^{125}$Te, $^{63}$Cu, and $^{65}$Cu nuclear spins with the applied external magnetic field and are only dominant at lowest temperature $T \leq 0.2\,\text{K}$ for the field range investigated here. Special care has to be taken for the Cu isotopes since the effective field at the position of the nuclei is a combination of the external field and the field due to the hyper-fine interaction of the electron spin and the nuclear spin, which can be determined from the zero field data for $T \leq 0.1\,\text{K}$ to $B_{e} \approx 10.2\,\text{T}$. For further details see Supplemental Information of Ref. \cite{Thurn2021}.

\section{Details of the data processing and evaluation of $\Gamma_{B}(B)$}
\label{AppendixB}
The experimental data of magnetic Gr\"uneisen parameter $\Gamma_{B}(B)$ were acquired via adiabatic field sweeps where the temperature of the sample $T(B)$ changes continuously as function of magnetic field $B$. $\Gamma_{B}(B)$ can then be directly calculated from the measured sample temperature via $\Gamma_{B}(B)=1/T\cdot(\partial T/\partial B)_{S}$. Since the sample temperature can easily double for a field sweep spanning the complete accessible field range, the influence of the temperature on $\Gamma_{B}(B)$ cannot be directly determined from such a measurement. Therefore multiple overlapping field sweeps over narrow field windows of $2 \,\text{T}$ width were combined to produce one single data set. This procedure enables $\Gamma_{B}(B)$ to be continuously measured within a temperature window of $50 \,\text{mK}$. 
To check for possible effects of non-adiabaticity, the results of these continuous measurements have been compared to results of discrete measurements of $\Gamma_{B}$ [see FIG.\,\ref{FieldExp}(a)]. A special situation was encountered for the field region $B > 11 \,\text{T}$ where $\Gamma_{B}(B)$ is quite large and a field change of $\Delta B = 1 \,\text{T}$ can lead to a temperature change of roughly $\Delta T = 150 \,\text{mK}$. Additionally, the large absolute value of the magnetic Gr\"uneisen parameter in the field region of $B > 11 \,\text{T}$ can produce a systematic error in the corresponding measurements of $\Gamma_{B}(B)$. For these measurements the large magnetic Gr\"uneisen parameter will generally lead to a rapid change of the sample temperature in a narrow field region which can in turn lead to a poor temperature control and therefore to deviations of ideal adiabatic conditions. In addition, eddy-current heating in the metallic parts of the sample holder has a stronger influence at lower temperature and high magnetic field where the specific heat of the sample is small and can thus cause artificial hysteresis loops between field-up and field-down measurements.\\

\section{Functional units of PbCuTe$_{2}$O$_{6}$ and their mutual interaction}
\label{AppendixC}
\begin{figure}[t!]
\centering
\includegraphics[width=0.47\textwidth]{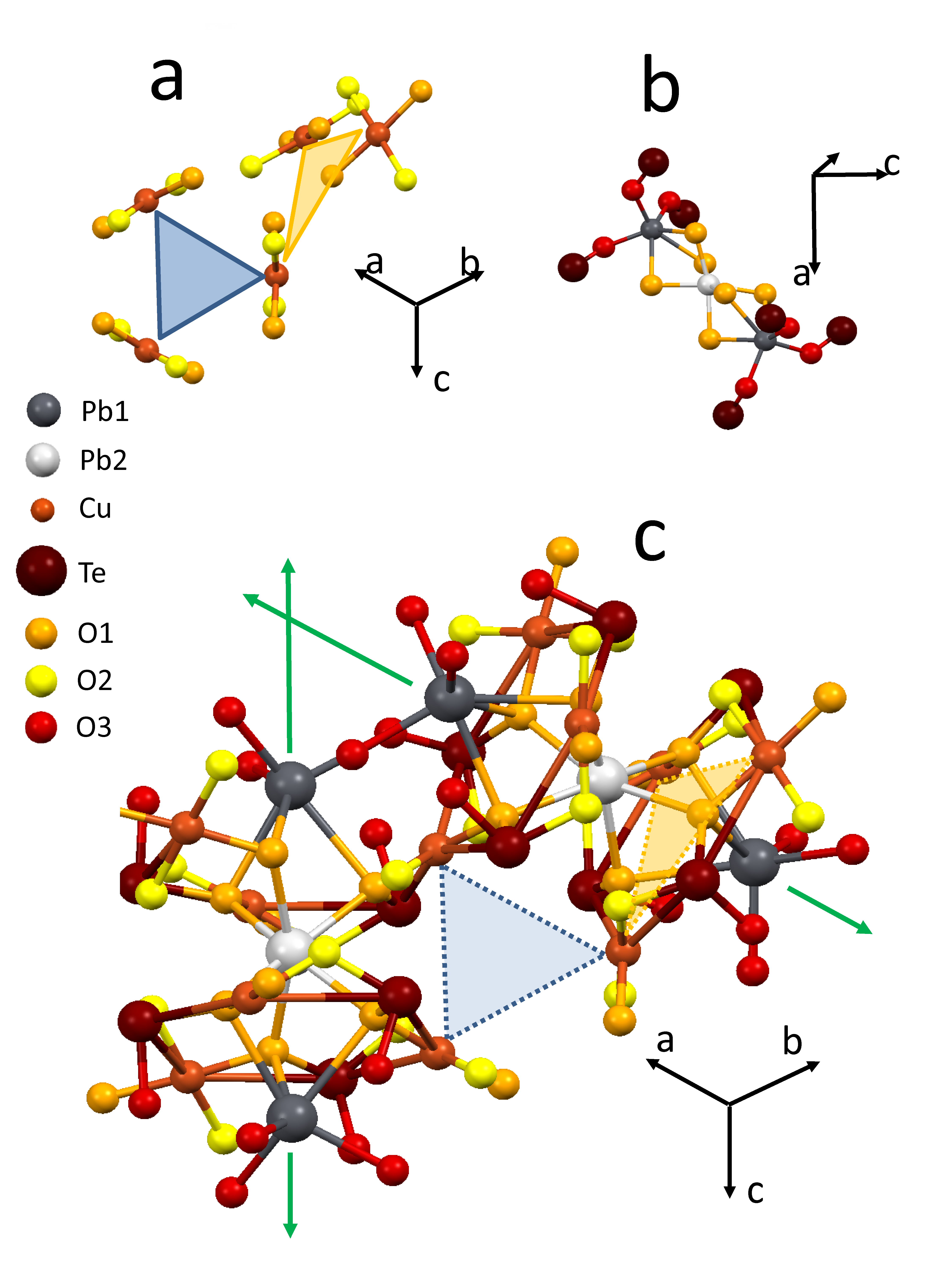}
\caption{The magnetic (a) and dielectric (b) functional units of the spin liquid candidate PbCuTe$_{2}$O$_{6}$ and their coupling in the lattice structure (c). For clarity several atoms are omitted in this representation (see text for more details). (a) View on selected magnetic units along the [111] direction. The planar oxygen-coordinated Cu$^{2+}$ ions have no direct orbital overlap with each other. The Cu$^{2+}$ ions are forming a highly frustrated hyperkagome lattice of edge-sharing triangles characterized by the two dominant coupling constants $J_1$ (blue triangle) and $J_2$ (yellow triangle). (b) The dielectric units are oriented along all 3-fold rotation axes of the non-centrosymmetric cubic structure. These units are built out of three Pb$^{2+}$ ions (two Pb1 ions and one Pb2 ion) on different crystallographic positions. The Pb2 ion (light grey) is symmetrically coordinated by six O1 ions. (c) Two adjacent dielectric units together with the planar oxygen-coordinated Cu$^{2+}$ ions along the [111] direction. The green arrows indicate 3-fold rotation axes and the blue and yellow triangles represent the nearest-neighbor [$J_1$ (blue)] and next-nearest-neighbor [$J_2$ (yellow)] magnetic interactions forming a 3D network of corner-sharing triangles known as a hyperkagome lattice. Both dielectric units are coupled with each other via the magnetic units by their O1 ions. The magnetic exchange path which leads to the coupling constant $J_1$ (blue triangle) is given by Cu-O1-Pb2-O1-Cu, whereas $J_2$ (yellow triangle) is given by Cu-O2-Te-O1-Cu.}
\label{FigXX}
\end{figure}

\indent
The spin liquid candidate PbCuTe$_{2}$O$_{6}$ features two types of functional units, magnetic- and dielectric ones. These functional units are forming a non-centrosymmetric cubic structure at high temperatures. The magnetic units are built out of planar fourfold oxygen-coordinated Cu$^{2+}$ ions with bond lengths of Cu-O1 = 1.911$\,\text{\AA}$ and Cu-O2 = 1.919\,$\,\text{\AA}$. These magnetic units form a highly frustrated hyperkagome lattice of edge-sharing triangles as indicated in FIG.\,\ref{FigXX}(a). The triangles, formed by three Cu$^{2+}$ ions at the corners, are characterized by the two dominant magnetic coupling constants $J_1$ (blue triangle) and $J_2$ (yellow triangle). Important for the understanding of the material's magnetic properties and the interaction between the magnetic and dielectric units is the fact that there is no direct orbital overlap between the planar oxygen-coordinated Cu$^{2+}$ ions. It was demonstrated in Ref.\,\cite{Liu2014} that Cu$^{2+}$ ions with $S$ = 1/2 in a square-planar coordination have a significant spin-orbit coupling which leads to a single-ion magnetic easy-plane anisotropy and a considerably strong spin-phonon interaction.\\
The dielectric units are oriented along all three-fold axes of the non-centrosymmetric cubic structure. These units are built out of (TeO$_3$)$^{2-}$ tetrahedra, where the Te atoms are covalently bonded to O1, O2 and O3 atoms, along with three Pb$^{2+}$ ions (two Pb1 and one Pb2) in different local coordinations. The (TeO$_3$)$^{2-}$ tetrahedra are expected to exhibit a permanent electric dipole moment also at high temperatures due to the lone pair of the Te$^{4+}$ valence electrons. These Te-derived electric dipoles are considered to be the main source of the enhanced dielectric constant reported in Ref.\,\cite{Thurn2021}. Further dielectric building blocks contain Pb$^{2+}$ ions and their oxygen coordination, displayed in FIG.\,\ref{FigXX}(b), consisting of two Pb1 ions (dark grey) and one Pb2 ion (light grey). The Pb1 ions are asymmetrically coordinated with three short bonds to the (TeO$_3$)$^{2-}$ tetrahedra via the O3 atoms with Pb1-O3 = 2.431$\,\text{\AA}$ and three long bonds between Pb1 and O1 of Pb1-O1 = 2.904$\,\text{\AA}$. In contrast, the Pb2 ion is symmetrically coordinated by six O1 ions with a bond length of Pb2-O1 = 2.531$\,\text{\AA}$. This means that these Pb2 ions have the potential to form stereochemically active lone pairs. This, however, requires the symmetry of their local environment to be reduced \cite{Walsh2011}. Based on the results of thermal expansion measurements [FIGs.\,\ref{FigLowBThermo} and \ref{FigExplHighB}], where we observe a structural instability to a non-cubic low-$T$ state with a precursor setting in below about 1.75\,K, we assume that these lattice distortions are such that they form an asymmetric Pb2-O1 coordination and thus a lone pair. The electric dipoles generated this way are considered to give rise to the observed ferroelectric order in this material around 1\,K \cite{Thurn2021}.
\\
To rationalize the interplay between the dielectric and the magnetic units, we show in FIG.\,\ref{FigXX}(c) two adjacent dielectric units together with the planar oxygen-coordinated Cu$^{2+}$ ions along the [111] direction. A common element shared by both the dielectric units and the magnetic units are the O1 ions. According to the above-sketched scenario, these O1 ions and their displacements are key for the ferroelectric order via the formation of stereochemically-active lone pairs. At the same time, the O1 ions are involved in the dominant magnetic exchange paths albeit in different ways. Whereas the magnetic exchange path giving rise to the coupling constant $J_1$ (blue triangle) is via Cu-O1-Pb2-O1-Cu, the coupling $J_2$ (yellow triangle) is mediated by Cu-O2-Te-O1-Cu. As a result, the structural distortions accompanying the formation of ferroelectric order are expected to change the ratio $J_1/J_2$, i.e., the degree of frustration, consistent with the observations made in Ref.\,\cite{Thurn2021}.

\end{appendix}
\pagebreak

\bibliography{PbCuTe2O6_finiteB_biblio}

\end{document}